\documentstyle{article}
\newcommand{\be}{\begin{equation}}
\newcommand{\ee}{\end{equation}}
\newcommand{\bea}{\begin{eqnarray}}
\newcommand{\eea}{\end{eqnarray}}
\newcommand{\nn}{\nonumber}
\newcommand{\hsp}{\hspace{3mm}}

\newcommand{\w}{\mbox{$\omega$}}
\newcommand{\eps}{\mbox{$\epsilon$}}
\newcommand{\vep}{\mbox{$\varepsilon$}}
\newcommand{\g}{\mbox{$\gamma$}}

\newcommand{\la}{\mbox{$\langle$}}
\newcommand{\ra}{\mbox{$\rangle$}}
\newcommand{\ad}{\mbox{$a^{\dag}$}}

\newcommand{\hb}{\mbox{$\hbar$}}

\begin{document}

\baselineskip 2em

\subsection*{On radiation reaction and the $[x,p]$ commutator
for an accelerating charge.}

\begin{center}
{\bf H. Fearn }\\
\small{Physics Department, California State University Fullerton\\
800 N. State College Blvd., Fullerton CA 92834 \\
{\bf email:} hfearn@fullerton.edu\\
{\bf phone:} (657) 278 2767}\\
{\bf url: } http://electron.fullerton.edu/$\sim $ neo \\

\vspace{0.25in}

\noindent
{\bf Abstract}\\
\end{center}

\noindent
We formally state the connection between the relativistic
part of the radiation reaction and the Poynting Robertson
force term, $-Rv/c^2$, where $R$ is power radiated.
Then we address the question, does $[x,p]=i \hb$ for an accelerating charge ?
The full radiation reaction term is used, which includes
the relativistic term (von Laue vector.).
We show that the full relativistic radiation reaction term
must be taken into account if a commutation relation between
$x$ and $p$ is to hold for an electron under uniform acceleration,
consistent with the expectation values of $x^2$ and $p^2$. \\

\noindent
{\bf PACS codes:} 41.60.-m,\hsp 41.75.Ht, \hsp 41.20.-q, \hsp  03.65.-w \\

\noindent
{\bf Key Words:} [ Commutation relations, accelerating electron,
radiation reaction, von Laue $\Gamma$, relativistic electron,
LAD equation, Poynting Robertson drag.]\\

\vspace{0.25in}

\section*{Introduction}

It has been shown \cite{mil}, that the nonrelativistic theory of the electron
is fundamentally inconsistent unless both radiation reaction and the
vacuum field are allowed to act on the electron. In particular, it was shown
that the canonical commutation rule $[x,p] = i\hb$ for the electron
is violated if radiation reaction is ignored. In this previous work \cite{mil},
the equation of motion used was that of Lorentz.
\be
\ddot{r}(t) - \g \ddot{r}^{\cdot}(t) = \frac{e}{m} E_0 (t)
\ee
where $\g = 2 e^2/3mc^3$ and $r$ is a vector.
We instead use the relativistic LAD equation of motion
(of Lorentz, Abraham and Dirac) \cite{lor,abr,dirac} given by
\be
\ddot{x}_{\nu} - \g (\ddot{x}_{\nu}^{\cdot} + \ddot{x}^{\mu} \ddot{x}_{\mu}
\dot{x}_{\nu}/c^2 ) = \frac{e}{m} \dot{x}_{\mu} F_{\nu}^{\mu}
\label{lad}
\ee
where the damping term $\g = 2 r_0/3c$ is defined above and has dimensions
of time,  and $r_0 = e^2/mc^2$ is the classical electron radius.
The term on the right is equivalent to $e E(t)/m$ where $E(t)$ is the electric
field.  Also $m$ is the rest mass, which we shall use throughout the paper.\\

Let us examine this equation (\ref{lad}) a little more closely.
The first term clearly comes from the kinetic energy of the
charged particle (electron in this case). The second term on the
left is the Abraham Lorentz radiation reaction term $F=2 e^2
\dot{a}/3 c^3$ or it is derived from what used to be called the
acceleration energy by Schott 1915 \cite{shot}. The third term on
the left is of the form $F=-Rv/c^2$ where $R=2e^2a^2/3c^3$ which
is the regular Larmor power radiated by an electron of
acceleration $a$. This term corresponds exactly to the
Poynting--Robertson force term \cite{poy,rob}. This is exactly the
Abraham result of 1903 \cite{abr} and the relativistic part of the
von Laue 1909 radiation reaction $\Gamma$ \cite{von}. A history of
the Dirac equation (\ref{lad}) is given by Rohrlich \cite{rohr}
and the book by Grandy \cite{grandy}. The remarkable thing about
Grandy's book is that on page 204 he writes the drag term (in the
LAD equation) in the format of the Poynting--Robertson drag
$-Rv/c^2$, where $R$ is the power radiated by the accelerating
charge. However, no connection is made between the (relativistic)
radiation reaction term and the Poynting--Robertson drag force.
The author feels that this is an oversight, no electromagnetic
text book mentions the connection even though the relativistic
term is very important and used all the time, possibly without
people realising it.  The two radiation reaction force terms are
both derived from the same power expression. Usually text books
(for example Jackson \cite{jack} or Griffiths \cite{griff}) do not
derive force expressions. They typically start with the electric
field of a moving charge, then calculate the Poynting vector and
then integrate over a cross--sectional area to find the power
radiated into a given angle.  Very few text books even mention the
famous Dirac paper of 1938. Jackson's book refers to it in the
problems section only, Griffiths' book does not mention it at all.
It should also be mentioned that Dirac's 1938 paper contains the
original discussion of the advanced and retarded fields,
which was elaborated on by Wheeler and Feynman later in 1945.\\

\noindent
When acceleration is taken into account, it is most important to
keep the relativistic radiation reaction term.
Boyer \cite{boy8}, noted that an electric dipole accelerated through the
vacuum would see a surrounding field not quite equal to the usual Planck
distribution. A correction term was needed \cite{boya},
which turned out to be exactly the relativistic radiation reaction term,
(or the Poynting and Robertson drag).
Boyer \cite{boya}, showed that a
\begin{quote}
`` classical electric dipole oscillator accelerating though
classical electromagnetic zero--point radiation responds just as would a
dipole oscillator in an inertial
frame in a classical thermal radiation with Planck's
spectrum at temperature $T=\hbar a/2 \pi c k$"
\end{quote}

\noindent
where $T$ is the Unruh--Davies temperature.\\

\noindent
Later, Boyer \cite{boyb}, did a similar calculation for the spinning
magnetic dipole and found a mismatch with
the Planck distribution again. He later corrected the magnetic
dipole work with a similar drag force to regain the Planck distribution,
\cite{boyc}. He was able to show that,
\begin{quote}
 `` the departure from Planckian form
is canceled by additional terms arising in the relativistic radiative damping
for the accelerating dipole. Thus the accelerating dipole behaves at
equilibrium as though in an inertial frame bathed by exactly Planck's spectrum
including zero--point radiation."
\end{quote}

\noindent
We have established that for an accelerated charge, consistency with
thermodynamics requires that we keep the relativistic radiation reaction
term in the equation of motion of the electron.
We now turn our attention to the $[x,p]$ calculation.
Instead of the quantized vacuum field, we will be dealing with the
vacuum field and the Planck distribution at a temperature given by the
Unruh Davies temperature $T= \hb a/(2 \pi c k_{\mbox{\tiny B}} )$ where
$a$ is the uniform acceleration and $k_{\mbox{\tiny B}}$ is Boltzmann's
constant \cite{boya,boy2}.\\

We are strongly motivated by Wigner \cite{wig}, who once asked the question;

\begin{quote}
``Do the equations of motion determine the quantum mechanical commutation
relations? "
\end{quote}
Wigner showed that the equations of motion for a harmonic oscillator do not
uniquely determine the commutation relations. We find that indeed the equation
of motion in our case does determine the commutation relation for $x$ and $p$.
The commutation relation found is consistent with the expectation values of
$x^2$ and $p^2$ but at first sight it appears it is not $[x,p] = i \hbar$. Our commutation relation
does however reduce to this expected result for the case when the acceleration
of the charge $a \rightarrow 0$.

\section*{Results and Discussion}

\noindent
Consider an electron undergoing an oscillation, like a
mass on a spring, with the motion restricted to the $x$ direction.
We shall take the resonant frequency of this oscillation to be $\w_0$.
This oscillatory motion is taken to be nonrelativistic and hence we continue to
use the rest mass of the electron $m$.
We consider the amplitude of this motion to be very small (as in the small
dipole approximation) and we only keep terms linear in the amplitude $x$ of the
oscillation. The electron is accelerating with uniform acceleration $a$ in
the $z$ direction.  We may rewrite the equation of motion in
this case as,
\be
\ddot{x} +\w_0^2 x - \g (\ddot{x}^{\cdot} - \frac{a^2}{c^2} \dot{x} ) =
\frac{e}{m} E(t)
\ee
where $E(t)$ is taken to be the electric field at the equilibrium point of
the oscillation, instantaneously at rest with respect to the oscillation,
as in Ref. \cite{boya}. Note
that we have ignored terms in the radiative damping that go like
$\ddot{x}^2 \dot{x}$ since these would involve the amplitude of the
oscillation squared which is a small quantity. We have only kept the
$z$--component of the acceleration squared term because this term is only
linear in the oscillation amplitude. We will write
the electric field, (in the rest frame of the equilibrium point of the
oscillation), in quantized form as
\be
E(t) = \frac{1}{\sqrt{\pi}} \int_0^{\infty} d\w \; \eps_{ks} \left(
\vep(\w_k ) a_{ks} e^{-i\w_k t} + \vep^{\star}(\w_k) \ad_{ks} e^{i \w_k t}
\right)
\ee
where $a_{ks}$ is the photon annihilation operator for mode $(\vec{k},s)$ and
$\eps_{ks}$ is the polarization unit vector. The Fourier components $\vep(\w)$
are well defined by Boyer \cite{boya} and in
earlier work by the same author cited therein. We leave the definition until
later. Note also that we have taken $1/\sqrt{\pi}$ and not
$1/\sqrt{ 2 \pi}$ since that would imply
an integral from $-\infty$ to $\infty$. Here we wish to keep the positive and
negative frequency components separate. Let
\be
x(t) = x_1 e^{-i \w_k t} + x_2 e^{i \w_k t}
\ee
and treat the positive and negative frequency terms separately in the equation
of motion.
Substituting $x_1$ into our Eq.(3) gives
\be
x_1 = -\frac{e}{\sqrt{\pi}m} \int_0^{\infty} d\w_k \; \frac{\eps_{ks}
\vep(\w_k) a_{ks}}{(\w_k^2 - \w_0^2) + i \g ( \w_k^3 + a^2 \w_k/c^2 )}
\ee
\noindent
similarly for $x_2$ we find
\be
x_2 = -\frac{e}{\sqrt{\pi}m} \int_0^{\infty} d\w_k \; \frac{\eps_{ks}
\vep^{\star}(\w_k) \ad_{ks}}{(\w_k^2 - \w_0^2 ) -
i \g ( \w_k^3 + a^2 \w_k/c^2 )}
\ee

\noindent
thus
\be
x(t) = -\frac{e}{\sqrt{\pi} m} \int_0^{\infty} d\w_k \; \eps_{ks} \left[
\frac{\vep(\w_k) a_{ks}e^{-i \w_k t}}{(\w_k^2 -\w_0^2) + i \g ( \w_k^3 +
a^2 \w_k/c^2 )}
+\frac{\vep^{\star}(\w_k) \ad_{ks}e^{i\w_k t}}{(\w_k^2 -\w_0^2 ) -
i \g ( \w_k^3 + a^2 \w_k/c^2 )} \right]
\ee
From this result we can easily find
\be
\dot{x}(t) = i\frac{e}{\sqrt{ \pi} m} \int_0^{\infty} d\w_k' \; \eps_{k's'}
\w_k' \left[ \frac{\vep(\w_k') a_{k's'}e^{-i \w_k' t}}{(\w_k'^2 -\w_0^2) +
i \g ( \w_k'^3 + a^2 \w_k'/c^2 )}
- \frac{\vep^{\star}(\w_k') \ad_{k's'}e^{i\w_k' t}}{(\w_k'^2 - \w_0^2 ) -
i \g ( \w_k'^3 + a^2 \w_k'/c^2 )} \right]
\ee

\noindent
The commutator then becomes
\bea
[x,p]&=&[x,m\dot{x}] \nn \\
&=& \frac{i e^2}{\pi m} \int_0^{\infty} d\w_k \; \int_0^{\infty} d\w_k'
\; \eps_{ks} \cdot \eps_{k's'} \vep(\w_k) \vep^{\star}(\w_k') \w_k' \left[
\frac{ }{ } \right. \nn \\
&& \frac{[ a_{ks},\ad_{k's'}] e^{-i(\w_k -\w_k') t}}{
[(\w_k^2 - \w_0^2 ) + i \g ( \w_k^3 + a^2 \w_k/c^2 )]
[(\w_k'^2 -\w_0^2 ) - i \g ( \w_k'^3 + a^2 \w_k'/c^2 )]} \nn \\
&& -\left. \frac{[ \ad_{ks},a_{k's'}] e^{i(\w_k -\w_k') t}}{
[(\w_k^2 -\w_0^2) - i \g (\w_k^3 +a^2 \w_k/c^2 )]
[(\w_k'^2 -\w_0^2) + i \g (\w_k'^3 +a^2 \w_k'/c^2 )]} \right]
\eea

\noindent
We have assumed here it is sufficient to use, $[a_{ks},\ad_{k's'}] = \delta_{ss'} \delta^3_{kk'}$ and
$[\ad_{ks},a_{k's'}] = -\delta_{ss'} \delta^3_{kk'}$ we find by setting
$\w_k = \w$ (we drop the $k$ subscript which is no longer needed),
\be
[x,p]= \frac{ie^2}{\pi m} \int_0^{\infty} d\w \; \frac{2 \w \la \vep(\w)
\vep^{\star}(\w) \ra }{(\w^2 -\w_0^2)^2 + \g^2 \w^6 [ 1 + (a/c\w )^2 ]^2 }
\label{xp}
\ee

\noindent
At this point we introduce the expectation of the fields given by Boyer
\cite{boya}, in Eq.(25) of that paper.
\bea
\la \vep(\w) \vep^{\star}(\w) \ra &=& \frac{4 \hb \w^3}{ 3 c^3}
\left[ 1 + \left( \frac{a}{c\w} \right)^2 \right]
\coth \left[ \frac{\pi c \w}{a} \right] \nn \\
&=& \frac{ 8 \pi^2 }{3c} \left[ 1 + \left( \frac{a}{c\w} \right)^2 \right]
\rho(\w) \coth \left[ \frac{\pi c \w}{a} \right]
\label{ee}
\eea

\noindent
where $\rho(\w) = \hb \w^3/(2 \pi^2 c^2 )$ the energy density of free space.
It is important to note at this stage that
\bea
\frac{1}{2} \coth \left[ \frac{\pi c \w}{a} \right] &=& \frac{1}{2}
\left( \frac{ e^{2\pi c\w /a} + 1}{ e^{2 \pi c \w /a} - 1} \right) \nn \\
&=&  \left( \frac{1}{2} + \frac{1}{e^{2 \pi c \w /a} -1 } \right)
\eea

\noindent
where $2 \pi c \w /a = \hb \w /( k_{\mbox{\tiny B}} T )$ which implies that
$T = \hb a/(2 \pi c k_{\mbox{\tiny B}} )$ which is the Unruh Davies temperature.
The $ \rho(\w) \coth $ term is therefore the vacuum field plus the Planck
distribution at temperature $T$.
Substituting the Eq.(\ref{ee}) into the integrand of Eq.(\ref{xp})
the commutator becomes,
\be
[x,p] = \frac{4 i \hb }{\pi} \int_0^{\infty} d\w
\frac{ \g \w^4 \left[ 1 + \left( \frac{a}{c\w} \right)^2 \right]
\coth \left[ \frac{\pi c \w}{a} \right] }{
(\w^2 -\w_0^2 )^2 + \g^2 \w^6 \left[ 1 +
\left( \frac{a}{c\w} \right)^2 \right]^2 }
\ee

\noindent
Then using the substitution $z = \w^2 - \w_0^2$ and $ s=a/(\w_0 c) $
and setting $\w = \w_0$ everywhere else we get,

\bea
[x,p] &=&  \frac{2 i \hb }{\pi} \int_0^{\infty} dz
\frac{ \g \w_0^3 (1+s^2) \coth \left[ \pi /s \right] }{z^2 +
\g^2 \w_0^6 (1+s^2)^2} \nn \\
&=&  \frac{2 i \hb }{\pi} \coth \left[ \pi /s \right] \int_0^{\infty} dz
\frac{A}{ z^2 + A^2 } \\
\eea
\noindent
where $A= \g \w_0^3 (1 + s^2)$. Hence,

\bea
[x,p] &=&  \frac{2 i \hb }{\pi} \coth \left[ \pi /s \right]
\left[ \tan^{-1} (z/A) \right]_0^{\infty} \nn  \\
&=& i \hb \coth \left[ \pi c \w_0 /a \right]
\eea

\noindent
where we have substituted back for $s$ in the last step.
This is not quite the expected result, but we do know that when
$a \rightarrow 0$ then $\coth[\infty]=1$ and we get back the expected result.
To show that this is indeed the correct result we need to consider the
uncertainty relation, $ \Delta x \Delta p  $.
Using,
\bea
\Delta x &=& \sqrt{ \la x^2 \ra  - \la x \ra^2 } \nn \\
\Delta p &=& m \sqrt{ \la \dot{x}^2 \ra  - \la \dot{x} \ra^2 } \nn \\
\eea
\noindent
and $ \la x \ra \equiv \la \dot{x} \ra =0$, using the results
from Boyer \cite{boya} we find;
\bea
\Delta x \Delta p &\geq & m \sqrt{ \la x^2 \ra \la \dot{x}^2 \ra } \nn \\
&\geq & \frac{\hbar}{2} \coth\left[\pi c \w_0 /a \right]
\eea

\noindent
Generally speaking, we know from quantum mechanics text books, that if we have
the uncertainty relation $\Delta a \Delta b \geq |d|/2$ then this
implies a commutator $[a,b] = id$. In our case above
$d = \hbar \coth\left[ \pi c \w_0 /a \right]$.\\

\noindent
\section*{Conclusions}

\noindent
In conclusion, we have made the connection between the relativistic
radiation reaction term (in the LAD equation of motion)
and the Poynting--Robertson drag force term.
We have further shown that it is necessary to use the full equation of motion
given by Lorentz, Abraham and Dirac (LAD) when treating an accelerated electron
in order that the commutation relation (which agrees with the
corresponding uncertainty relation) between $x$ and $p$ hold.
The result reduces to the expected result when the acceleration
$a \rightarrow 0$.\\

We also note that we have assumed a form of electric field
quantization without a rigorous derivation. Our end result does
appear consistent with previous work. However, we note that to
regain the regular commutation relation it may be necessary to
consider the relativistic Dirac equation, and use a combination of
$\alpha$ or $ \gamma$ matrices rather than $x$ and $p$ operators.

Finally, we note that is it possible to derive the result quickly by using the vacuum thermofield
operators originally defined by Takahashi and Umezawa \cite{taka,fearn} for the $a$ and $\ad$ operators
which then gives the $\coth\left[ \pi c \w_0 /a \right]$ term directly in the commutator.
The thermofield commutator for
$[a_{ks},\ad_{k's'}]_T = \coth\left[ \pi c \w_0 / a \right] \delta_{ss'} \delta^3_{kk'}$
would replace the commutator above Eq.(11), where we have used the Unruh Davies temperature $T$. \\

\noindent
The Thermofield operators use a vacuum double Hilbert space, with a fictitious mode or virtual photon mode
for the vacuum. The ordinary $a$ and $\ad$ operators are transformed
via a Bogoliubov transformation  from a unitary operator $T(\theta)$. The thermofield vacuum is given by,
\be
| 0 \ra_T = T(\theta) | 0 \tilde{0} \ra
\ee
where  $T(\theta) = \exp[ -\theta ( a \tilde{a} - \ad \tilde{a}^\dag )]$, where the
tilde denotes a fictitious or virtual mode.
Note that $ T^\dag (\theta) = T (-\theta ) $. Using
\be
 e^A B e^{-A} = B + [A,B] + \frac{1}{2!} [A,[A,B]] + \frac{1}{3!} [A,[A,[A,B]]] + ... , \nn \\
\ee
we find,
\bea
a_T = T(\theta) a T^\dag (\theta ) &=& a \cosh \theta - \tilde{a}^\dag \sinh \theta \nn \\
{a^\dag}_T = T(\theta) \ad T^\dag (\theta ) &=& \ad \cosh \theta - \tilde{a} \sinh \theta .\\
\eea
It is easy to show that the photon number and commutator for the thermofield operators become,
\bea
\la 0 \tilde{0} | \ad_T a_T | 0 \tilde{0} \ra &=& \sinh^2 \theta = \frac{ 1}{ e^{2\alpha} -1 } \nn \\
\la 0 \tilde{0} |[ a_T , \ad_T ] | 0 \tilde{0} \ra &=& \sinh^2 \theta + \cosh^2 \theta = \coth \alpha \\
\eea
where $ \alpha = \pi \w_0 c /a $. Clearly we could have used these for our accelerating frame
operators and saved a bit of time calculating electric fields like Boyer \cite{boya}. It is not clear
if the thermofield operation can be related to space-time curvature, and quantum gravity,
but the accelerated frame appears to give the same result as would the thermal frame using
thermofield operators and the Unruh Davies temperature relation. It may be that we should simply
 normalize the $[a,\ad]$ commutator to account for the thermal result,
 $\coth\left[ \pi c \w_0 / a \right]$, in which case you regain the usual $[x,p]$ commutator. This
 would also normalize the uncertainty relation and we would get back $\Delta x \Delta p \geq  \hbar/2 $.

 \noindent
 \subsection*{Acknowledgements}

 \noindent
 This work has been in preparation for a long while. Most of the theory was completed while HF
 was at LANL over 10 years ago and parts have been used in lecture notes given in classes at
 CSU Fullerton Physics department. HF wishes to thank the physics department for a sabbatical
 leave in the Fall 2011 when the manuscript was mostly completed.


\newpage


\begin{thebibliography}{9}

\bibitem{mil} P. W. Milonni,  ``Radiation reaction and the nonrelativistic
theory of the electron", Phys. Letts. {\bf 82A}, pp225--226 (1981).

\bibitem{lor} H. A. Lorentz, ``The Theory of Electrons", (Dover Pubs. Inc.
New York, 1904) reprinted 1915.

\bibitem{abr} M. Abraham, Annalen der Phys. {\bf 10}, pp105--179 (1903).

\bibitem{dirac} P. A. M. Dirac, Proc. Roy. Soc. Lon. {\bf A167}, pp148--169
(1938).

\bibitem{shot} G. A. Schott, Phil. Mag. {\bf 29}, pp49--62 (1915).

\bibitem{poy} J. H. Poynting, Phil. Trans. Roy. Soc. {\bf A202}, p525 (1903),
reprinted with corrections in his ``Collected Scientific Papers" p304 (1920).

\bibitem{rob} H. P. Robertson, Monthly Notices of the Royal Astron. Soc.
{\bf 97} pp 423--438 (1937). See also H. P. Robertson and T. W. Noonan,
``Relativity and Cosmology", (W. B. Saunders Company, Philadelphia, London,
Toronto 1968).

\bibitem{von} M. von Laue, Annalen der Phys. {\bf 28}, pp436--442 (1909).

\bibitem{rohr} F. Rohrlich, Am. J. Phys. {\bf 65}, pp1051--1056 (1997) and
Am. J. Phys. {\bf 68}, pp1109--1112 (2000).

\bibitem{grandy} W. T. Grandy, ``Relativistic quantum mechanics of leptons
and fields", (Kluwer Academic Press, Dordrecht, Boston and London 1991). See
Chapter 7 page 204.

\bibitem{jack} J. D. Jackson ``Classical Electrodynamics", third edition
(John Wiley \& Sons, Inc, New York 1999).

\bibitem{griff} D. J. Griffiths, ``Introduction to Electrodynamics", third
edition, (Prentice Hall, New Jersey 1999).

\bibitem{boy8} T. H. Boyer, Phys. Rev. {\bf D21}, pp2137--2148 (1980).

\bibitem{boya}  T. H. Boyer, Phys. Rev. {\bf D29}, pp1089--1095 (1980).

\bibitem{boyb} T. H. Boyer, Phys. Rev. {\bf A29}, pp2389--2394 (1984).

\bibitem{boyc} T. H. Boyer, Phys. Rev. {\bf D30}, pp1228--1232 (1984).

\bibitem{boy2} T. H. Boyer, Phys. Rev. {\bf D11}, pp809--830 (1975).

\bibitem{wig} E. P. Wigner, Phys. Rev. {\bf 77}, 711 (1950).

\bibitem{taka} Y. Takahashi \& H. Umezawa Coll. Phenom. {\bf 2}, 55 (1975).

\bibitem{fearn} H. Fearn, Quantum Optics {\bf 2}, pp103--118 (1990).

\end{thebibliography}
\end{document}